\documentclass[acmlarge,screen]{acmart} 

\usepackage{enumitem}
\usepackage{mdframed}

\newcommand{\italquote}[2]{\begin{quote}\textit{#1}~(#2)\end{quote}}
\newcommand{\revision}[1]{\textcolor{black}{#1}}

\AtBeginDocument{%
  }

\setcopyright{acmlicensed}
\acmJournal{IMWUT}
\acmYear{2024} \acmVolume{8} \acmNumber{3} \acmArticle{116} \acmMonth{9}\acmDOI{10.1145/3678575}

\begin{document}

\title{Users' Perspectives on Multimodal Menstrual Tracking Using Consumer Health Devices}

\author{Georgianna Lin}
\orcid{0000-0002-9993-2718}
\affiliation{%
  \institution{University of Toronto}
  \country{Canada}
  }
\email{blue.lin@mail.utoronto.ca}

\author{Brenna Li}
\orcid{0000-0003-3692-243X}
\affiliation{%
  \institution{University of Toronto}
  \country{Canada}
  }
\email{Brli@cs.Toronto.edu}

\author{Jin Yi Li}
\orcid{0000-0003-4100-8327}
\email{helenjy.li@mail.utoronto.ca}
\affiliation{%
  \institution{University of Toronto}
  \country{Canada}
 }

\author{Chloe Zhao}
\orcid{0000-0002-3087-9285}
\affiliation{%
  \institution{University of Toronto}
  \country{Canada}
  }
\email{yichen.zhao@mail.utoronto.ca}

\author{Khai Truong}
\orcid{0000-0003-0774-5964}
\affiliation{%
  \institution{University of Toronto}
  \country{Canada}
  }
\email{khai@cs.toronto.edu}

\author{Alexander Mariakakis}
\orcid{0000-0002-9986-3345}
\affiliation{%
  \institution{University of Toronto}
  \country{Canada}
  }
\email{mariakakis@cs.toronto.edu}

\renewcommand{\shortauthors}{Lin et al.}

\begin{abstract}
Previous menstrual health literature highlights a variety of signals not included in existing menstrual trackers because they are either difficult to gather or are not typically associated with menstrual health. 
Since it has become increasingly convenient to collect biomarkers through wearables and other consumer-grade devices, our work examines how people incorporate unconventional signals (e.g., blood glucose levels, heart rate) into their understanding of menstrual health.
In this paper, we describe a three-month-long study on fifty participants’ experiences as they tracked their health using physiological sensors and daily diaries. 
We analyzed their experiences with both conventional and unconventional menstrual health signals through surveys and interviews conducted throughout the study.
We delve into the various aspects of menstrual health that participants sought to affirm using unconventional signals, explore how these signals influenced their daily behaviors, and examine how multimodal menstrual tracking expanded their scope of menstrual health.
Finally, we provide design recommendations for future multimodal menstrual trackers.
\end{abstract}




\maketitle

\section{Introduction}
\label{sec:intro}
Rising gendered healthcare costs~\cite{glynn2015taming}, restrictions in relevant medical services~\cite{nobles2022menstrual}, and stigmatization~\cite{johnstonrobledo2011menstrual} place economic and social pressure on individuals to seek out medical information regarding their gendered health needs on their own \cite{lin2022investigating, epstein2017examining}.
Many individuals interested in learning more about their menstruation turn to period-tracking smartphone apps that account for basic inputs such as flow timing and duration~\cite{zwingerman2020critical}. 
Unfortunately, these data streams alone are often not reflective of the unique characteristics of users who menstruate and those who are transitioning into and out of menstruation (e.g., due to menarche, pregnancy, or menopause)~\cite{epstein2017examining, fox2020monitoring, keyes2020reimagining, kumar2020taking, lin2022investigating, zwingerman2020critical}.

\citet{hennegan2021menstrual} argue that menstrual health encompasses “a state of complete physical, mental, and social well-being and not merely the absence of disease or infirmity in relation to the menstrual cycle.”
Clinical literature has also supported the relevance of various signals to menstrual health.
Variations in heart rate~\cite{brar2015effect}, peripheral body temperature, sleep duration~\cite{driver1998menstrual}, and blood glucose~\cite{ykijarvinenf1984insulin, lin2023blood} across the menstrual cycle influence and can be influenced by an individual's dietary and exercise practices, sleep habits, and cardiovascular health.

The ability to capture multimodal data relevant to menstrual health has been made possible by wearables and other consumer-grade technologies that can passively collect heart rate and body temperature data~\cite{maijala2019nocturnal, shilaih2018modern, shilaih2017pulse}. 
However, many individuals who menstruate and the period-tracking apps they have embraced have yet to integrate these signals into their scope of menstrual health.
Conversely, multimodal tracking has become more prominent across health topics ranging from physical activity to stress management~\cite{amin2016on, choe2017understanding, ferreira2015aware, kim2017omnitrack, banos2016human, wu2021multimodal, radu2018multimodal}, yet menstrual health has been largely neglected.
The lack of multimodal tracking tools that consider menstrual health in any capacity potentially reflects the persistence of stigma and lack of holistic perspectives on menstrual health~\cite{johnstonrobledo2011menstrual,babbar2022menstrual,kumar2020taking,lin2022investigating}.

Our work seeks to identify the opportunities and challenges associated with using ubiquitous technologies to expand people's purview of menstrual health.
To explore these considerations, we investigate the following research questions:
\begin{enumerate}
    \item[\textbf{RQ1.}] What preconceptions do people have about menstrual health that they seek to confirm while engaging in multimodal menstrual tracking?
    \item[\textbf{RQ2.}] In what ways does multimodal menstrual tracking change people's menstrual health routines?
    \item[\textbf{RQ3.}] In what ways does multimodal menstrual tracking change people's understanding of their menstrual health?
\end{enumerate}

To answer these questions, we recruited fifty individuals who menstruate to engage in diverse health-tracking practices over three months.
Participants were asked to track information that varied along two dimensions: (1) the availability of the data to the general public, and (2) the degree to which people recognized the connection between the corresponding data and their menstrual health.
For the purposes of our work, we consider menstrual health signals that are both widely available and recognized within menstrual health contexts to be \textit{conventional} and signals that lack in either of these qualities to be \textit{unconventional}.
Participants used a daily diary to record characteristics of their menstrual health and replicate conventional menstrual tracking practices.
To capture unconventional menstrual health data, they used an at-home hormone analyzer daily to examine a signal that most of them knew was associated with their menstrual health but did not have the opportunity to collect.
They also wore two wearables to continuously monitor signals that were less commonly associated with menstrual health: a commonplace wrist-worn fitness tracker and a less prevalent continuous glucose monitor.
Participants reported their experiences via surveys and interviews administered throughout the study.
We use the phrase \textit{multimodal menstrual tracking} to describe tracking behaviors that involve collecting both conventional and unconventional menstrual health data simultaneously to understand the relationships between them.

Our findings highlight the ways that unconventional signals can reshape conventional menstrual health routines and understandings through multimodal menstrual tracking.
Participants were able to explore their menstrual health across various menstrual cycle phases beyond menstruation, which helped draw tighter connections between their menstrual cycles and associated symptoms.
Participants also commented on the potential benefits of incorporating unconventional menstrual health signals into menstrual and universal health trackers, such as building positive associations with menstrual health and setting nuanced expectations for menstrual cycle regularity.
These findings inspired us to generate a number of design implications for future multimodal menstrual trackers.
For example, we argue that more fine-grained menstrual phase predictions could help users identify correlations within and outside of menstrual contexts.
We also suggest affordances that help users understand reasonable fluctuations in physiological data over multiple cycles while alleviating concerns over intra- and inter-personal differences.
By expanding the scope of menstrual trackers and the data they collect, we envision tools that can capture and convey the range of menstrual health experiences over time and across individuals according to both conventional and unconventional menstrual health data, a task currently achievable only through daily clinical visits.
Since more continuous data can lead to increased learning and better prediction accuracy, our work prompts further discussions on how gendered health data is handled given the potential misuse of stigmatized data against individuals' wishes and autonomy.

\section{Related Work}
In our review of prior work, we first explain how health informatics research has embraced gendered health topics.
We then describe emerging methods of ubiquitous tracking that have been motivated by these efforts.

\subsection{Menstrual Health Informatics}
\label{sec:rwmhinfo}
Health informatics is a rich field that studies how people collect, understand, and share personal health data.
\citet{epstein2015lived} enumerate the varied motivations that people have for health tracking, including behavior change, activity instrumentation, and curiosity.
These motivations vary across domains, influencing how individuals select and employ tracking methods.
Regarding gendered health, there have been many targeted efforts to explore the needs of those transitioning in and out of menstruation.
Relevant subjects in this area have included maternal health~\cite{balaam2015feedfinder, d2016feminist}, reproductive health~\cite{chandler2021developing, chandler2020promoting}, childhood development~\cite{kientz2009baby, suh2014babysteps}, and menopause~\cite{bardzell2019re, lazar2019parting, tutia2019hci, homewood2019inaction}. 
More broadly, researchers have explored general health concerns within the context of menstrual health: health literacy~\cite{eschler2019defining, jarrahi2020comparison}, health inequities~\cite{kaufman2005china, bui2012dimensions, elmore2021health, samarasekera2017women, brown2020black, lin2022investigating}, health stigmas~\cite{tuli2018learning, tuli2019sa, tuli2020menstrual}, intimate bodily knowledge~\cite{campo2020touching, almeida2016hci, almeida2016looking}, wellbeing~\cite{almeida2016hci, kannabiran2018design, kumar2020taking}, and mental health~\cite{doherty2019engagement}.

At the intersection of gendered health and health informatics, technologies have been proposed to predict menstrual health information such as the date, duration, and health of a user's menstrual and ovulation cycles~\cite{epstein2017examining}.
Scholars have noted various shortcomings of these tools, one of the most prominent being the inaccuracy of the information and predictions they present~\cite{epstein2017examining,zwingerman2020critical,fox2020monitoring}.
\citet{zwingerman2020critical} partly attribute this issue to the fact that fewer than half of the apps in their review provide predictions based on user-reported data over time. 

Another set of problems that have been identified involves the lack of inclusivity in technology designs.
\citet{epstein2017examining} found that menstrual trackers often stereotype and are inaccessible, while \citet{fox2020monitoring} found that some menstrual trackers include heterosexist iconography and ageist resources.
\citet{lin2022investigating} extended this literature by investigating how menstrual trackers fail to support those with minimal menstrual health education and those who are not in the sexual majority.
\citet{tuli2022rethinking} argue that this problem is complicated by the varied stigmas that individuals who menstruate experience while going through different stages of their menstrual journey (e.g., menarche, pregnancy, menstruation).
Underlying all of these shortcomings is the perpetuation of discriminatory and non-reflexive notions of “universal womanhood” throughout existing tracking methods~\cite{keyes2020reimagining, kumar2020taking, eschler2019defining, lin2022investigating, bardzell2010feminist, epstein2017examining, zwingerman2020critical, fox2020monitoring}.

In light of these issues, \citet{kumar2020taking} appeal for a broader definition of gendered health that encompasses an intersectional view of health and wellbeing rather than one that focuses specifically on maternal, sexual, and reproductive health. 
\citet{figueiredo2021goals} also call for comprehensive health tracking that accompanies users throughout their various life stages.
Clinical literature has already acknowledged many relationships between aspects of menstrual health and health broadly construed.
For example, sports science research has identified interwoven physiological relationships between exercise and menstrual cycles.
Hormones like estrogen and progesterone impact both the menstrual cycle and exercise performance~\cite{oosthuyse2010effect}; simultaneously, exercise also influences menstrual cycle symptoms and mood states~\cite{aganoff1994aerobic}.
Other examples of bi-directional relationships with menstrual health entail heart rate variability~\cite{brar2015effect}, insulin~\cite{ykijarvinenf1984insulin}, sleep~\cite{driver1998menstrual}, and stress~\cite{sommer1978stress}. 

Incorporating such measures within current menstrual trackers would require not only the ability to either manually input data or integrate an external data stream, but also the intelligence to help users draw connections between the various data sources.
However, \citet{pichon2021messiness} find that current trackers often lack support for tracking unconventional menstrual health signals and instead reinforce a stereotype of relevant menstrual signals (e.g., cramps, mood swings)~\cite{wongkhoo2010}.
Building on findings from prior research, we explore how users respond to menstrual health signals that vary along two dimensions: (1) the availability of the signal and the technologies that can record it, and (2) the degree to which people recognize the connection between the signal and their menstrual health.

\subsection{Ubiquitous Technologies for Health Tracking}
\label{sec:rwubi}
Within the realm of health informatics, unobtrusive technologies like wearables and other consumer-grade devices have demonstrated the potential to positively influence users' behaviors by showing them “meaningful data abstractions and intuitive feedback mechanisms”~\cite{ananthanarayan2012persuasive}.
For example, the OmniTrack mobile system offered users the flexibility to create their own customized multimodal trackers to track data according to their individual needs and preferences~\cite{kim2017omnitrack}.
The Health Mashups mobile application~\cite{bentley2013health} and the Visualized Self~\cite{choe2017understanding} web application also incorporated various data sources into one centralized tracker to help users identify associations across multimodal data streams.
Observing users of the aforementioned systems led to some common findings: visualizations of and interactions with multimodal data foster a greater sense of introspection and self-reflection~\cite{bentley2013health,choe2017understanding, kim2017omnitrack,choe2014understanding}, data-driven observations allow users to prioritize changing certain behaviors~\cite{bentley2013health,choe2017understanding,choe2014understanding}, and users often voluntarily share their data with family and friends~\cite{rooksby2014personal,kim2017omnitrack}.
However, none of these systems accounted for data that would typically be recorded in current menstrual trackers, nor were they catered to questions that individuals would seek to answer about menstrual health.

Although most ubiquitous devices are marketed to the general public, some have been used for investigations specifically within menstrual contexts.
For example, \citet{wave} proposed designs for a smart-mirror that can display menstrual health data, while \citet{maijala2019nocturnal} explored the relationship between finger skin temperature and menstrual cycle timing using the Oura ring\footnote{\url{https://ouraring.com/}} data.
Most work involving ubiquitous devices for gendered health has only focused on comparing collected data with ground-truth signals~\cite{Zhu2021TheAO, nulty2022ava, maijala2019nocturnal}, supporting traditional clinical fertility services~\cite{goodale2021wreal}, and tracking physiological symptoms associated with fertility~\cite{shilaih2017wearable, shilaih2018wearable, goodale2019wearable, hayano2021prediction}. 
In other words, these works have not considered the varied reasons that people choose to track their menstrual health, including self-reflection and expanding menstrual literacy.

Rather than creating bespoke health tracking tools specifically for menstrual health, one could consider adapting multimodal health tracking systems specifically for menstrual health.
However, prior studies in personal informatics have highlighted that data may lose meaning when it is removed from its context, whether that context be defined by routines (e.g., workouts, commute), special events (e.g., marriage, holiday, moving), or systematic changes (e.g., seasons, weather)~\cite{amin2016on,choe2017understanding,ferreira2015aware,kim2017omnitrack,banos2016human,wu2021multimodal,radu2018multimodal}.
While work is being done to incorporate external context into multimodal tracking~\cite{bentley2013health,rooksby2014personal,kim2017omnitrack,choe2017understanding,choe2014understanding}, menstrual contexts have generally been neglected when it comes to interpreting and reading health data from these tracking practices, despite the fact that a worldwide majority experiences decades of menstruation~\cite{brantelid2014menstruation}.
This negligence may be due to stigmatization of the menstrual cycle or a lack of awareness about the relevance of unconventional menstrual health signals~\cite{diaz2006menstruation}.

As noted by \citet{li2011understanding}, designing tools that assist in self-reflection requires understanding the types of questions people have about their data, the reasons behind these questions, how people currently address these questions using available tools, and the challenges they encounter in the process.
Our work extends prior literature by studying how individuals who menstruate react when they use existing multimodal tracking methods to reflect on their health in the menstrual context.

\section{Methods}
In this section, we describe how we recruited individuals interested in menstrual tracking.
We then report the protocol they followed and the methods we used to analyze their experiences.
Building on the methodology of \citet{moore2021interview}, our study delved into the longitudinal interactions that participants had with their data.
We aimed to gain insight into how people's menstrual-tracking motivations and behaviors evolve over time, and we also aimed to understand the questions and discoveries participants generated from their data.
This study protocol was approved by the Research Ethics Board at the University of Toronto under Protocol \#41568.

\subsection{Participants}
\begin{table*}[!t]
    \caption{The demographics of our study population (N = 50).}
    \label{tab:methods-demographics}
    \Description{The demographics of our fifty participants are shown here. Age ranges from 18 to 29. Participants come from a plethora of racial and educational backgrounds and mostly identify as women. Participants also identified varied menstrual experiences, menstrual education levels, and menstrual tracking experiences.}
    \begin{tabular}{ l c }
    \toprule
      \textbf{Age (years)} & Min = 18, Max = 29, Mean = 20.6, Median = 20\\ \hline
      \textbf{Race} & East Asian (15), South Asian (9), SE Asian (3), Caucasian (14)\\ & Middle Eastern (4), African (3), Latina (1), Indo-Caribbean (1)\\ \hline
      \textbf{Self-Identified Gender} & \revision{Woman} (44), Gender-fluid (1), Non-binary (3)\\ & Questioning (1), Unanswered (1)\\ \hline
      \textbf{Self-Identified Menstrual Experience} & Regular \revision{(36)}, Irregular \revision{(14)}\\ \hline
      \textbf{Self-Identified Menstrual Education Level} & Non-existent (1), Low (5), Medium (30), High (13), Expert (1)\\ \hline
      \textbf{Education} & High school graduate, equivalent, or less (2)\\ & Some university / post-secondary, no degree (23), \\ & Bachelor's (10), Master's (1), Doctorate or professional degree (4)\\ &Unanswered (10)\\ \hline
      \textbf{Menstrual Tracking Experience} & Currently tracking (37), Stopped tracking (5), Never tracked (8)\\ \hline
      \textbf{Current Menstrual Tracker} & Clue (18), Flo (10), Apple Calendar (8), Fitbit (1), \revision{None (13)}\\
    \bottomrule
    \end{tabular}
\end{table*}
\label{sec:participants}

We enrolled 50 participants by posting advertisements on social media groups and workspaces operated by health advocacy organizations in the Greater Toronto Area.
We also leveraged email chains within the university and the authors' social networks to reach a wider audience.
Recruitment was limited to individuals who menstruate over the age of 18 who were not taking hormonal therapy or contraception for at least three months prior to the study.
Although there was not a maximum age restriction, we focused our efforts on recruiting participants who did not anticipate entering perimenopause or menopause as this transition can have significantly different needs and expectations~\cite{mckinlaymenopause1992}.
The demographic information of the participants is summarized in \autoref{tab:methods-demographics}.

\subsection{Study Design}
\label{sec:study-design}
\begin{figure*}
    \centering
    \includegraphics[width = 0.8\textwidth]{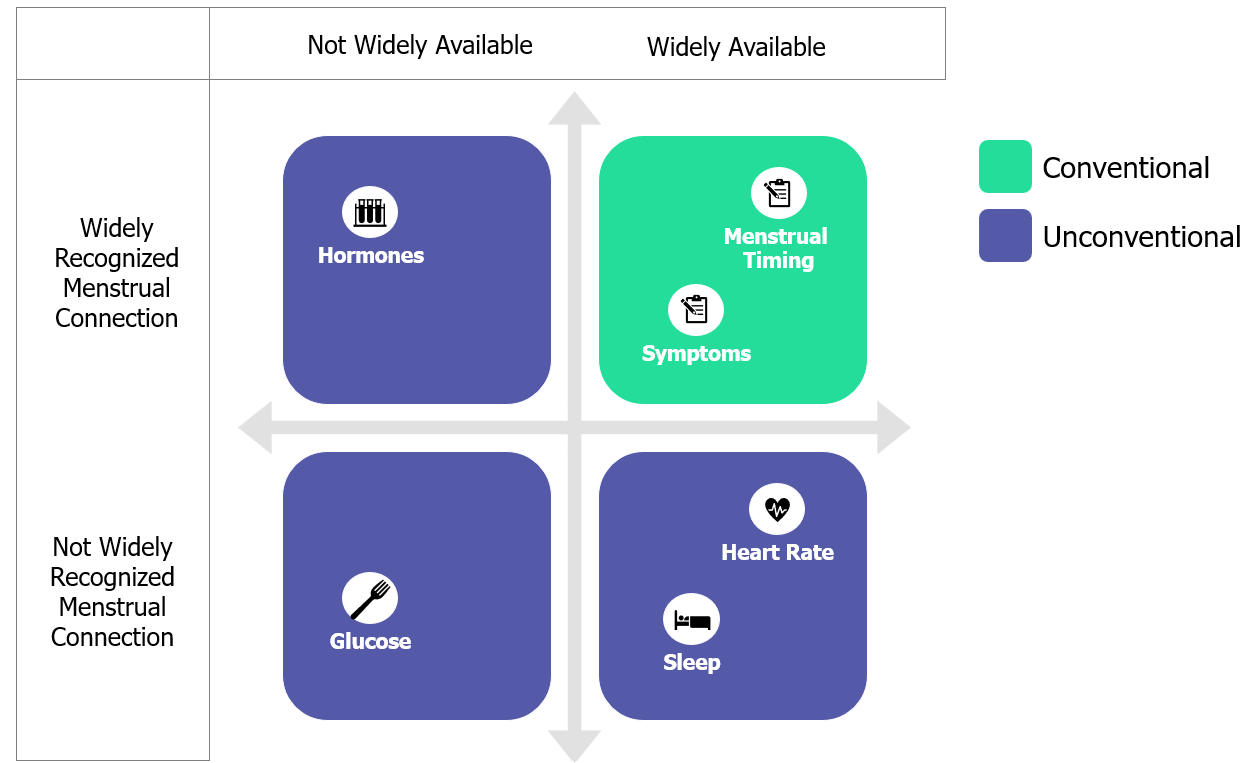}
    \caption{
    Participants engaged with a diverse range of data collection methods that differed in their availability and the degree to which people recognized the connection between the corresponding data and their menstrual health. 
    We use these two dimensions to differentiate conventional and unconventional menstrual health signals.
    }
    \label{fig:axesmenstrual}
    \Description{Participants engaged with a diverse range of data collection methods for multimodal menstrual tracking that differed in their availability and the degree to which people recognized the connection between the corresponding data and their menstrual health. We used these two dimensions to differentiate conventional and unconventional menstrual health signals. Menstrual timing and self-reported symptoms are shown as a widely available and conventional signal that is widely recognized as being connected to menstrual health. All other dimensions are considered unconventional. For example, hormones are shown as not widely available but widely recognized menstrual connections. Heart rate and sleep are shown as widely available and not recognized menstrual connections. Finally, glucose is shown as both not available and not recognized menstrual connection.}
\end{figure*}

To inform the design of our study protocol, we reviewed clinical literature to identify health signals associated with menstrual health and cross-referenced them with available market devices capable of tracking menstrual-related signals. 
This process led us to select the following data collection methods:
\begin{enumerate}
    \item \textbf{Hormone Analyzer:} Participants were asked to use a Mira Plus Starter Kit\footnote{\url{https://usd.miracare.com/products/fertility-plus-starter-kit}} daily to track their luteinizing hormone (LH) and estrogen (E3G) levels.
    \item \textbf{Fitness Tracker:} Participants were asked to wear a Fitbit Sense smartwatch\footnote{\url{https://www.fitbit.com/global/us/products/smartwatches/sense}} to record their vital signs and proprietary metrics related to their physical activity~\cite{oosthuyse2010effect}, sleep~\cite{driver1998menstrual}, and mindfulness~\cite{sommer1978stress}.
    \item \textbf{Glucose Monitor:} Participants were asked to wear a Dexcom G6 continuous glucose monitor\footnote{\url{https://www.dexcom.com/g6-cgm-system}} (CGM) to measure their glucose level~\cite{ykijarvinenf1984insulin, lin2023blood}.
    \item \textbf{Daily Diary:} Participants were asked to record the timing of their menstruation and the perceived severity of their menstrual symptoms in order to emulate the standard approach to menstrual tracking~\cite{epstein2017examining}.
\end{enumerate}

These data collection methods and health signals vary along two dimensions that were relevant to our research aims: (1) the availability of the signal to the general public, and (2) the degree to which the general public recognizes the connection between the signal and their menstrual health.
We consider health signals that are both readily available and typically associated with menstrual health to be conventional menstrual health signals, while those lacking in either dimension could be considered unconventional.
We consider multimodal menstrual tracking to be the act of collecting both conventional and unconventional menstrual health data for the purpose of understanding menstrual health.

\autoref{fig:axesmenstrual} illustrates where the data collection methods employed in our study reside within this taxonomy.
Current menstrual trackers typically fall favorably along both dimensions, relying on self-reported health signals clearly associated with menstrual health.
Hormone analyzers and continuous glucose monitors are less pervasive, making them unconventional according to the availability axis.
Meanwhile, data from fitness trackers and glucose monitors are not as commonly associated with menstrual health, making them unconventional according to the other dimension.
We note that the placement of data collection methods along this taxonomy is not rigid, as a signal may be more accessible or relevant to menstrual health for some individuals than others.

\subsection{Procedure}
\label{sec:procedure}

\begin{figure*}
    \centering
    \includegraphics[width = 0.95\textwidth]{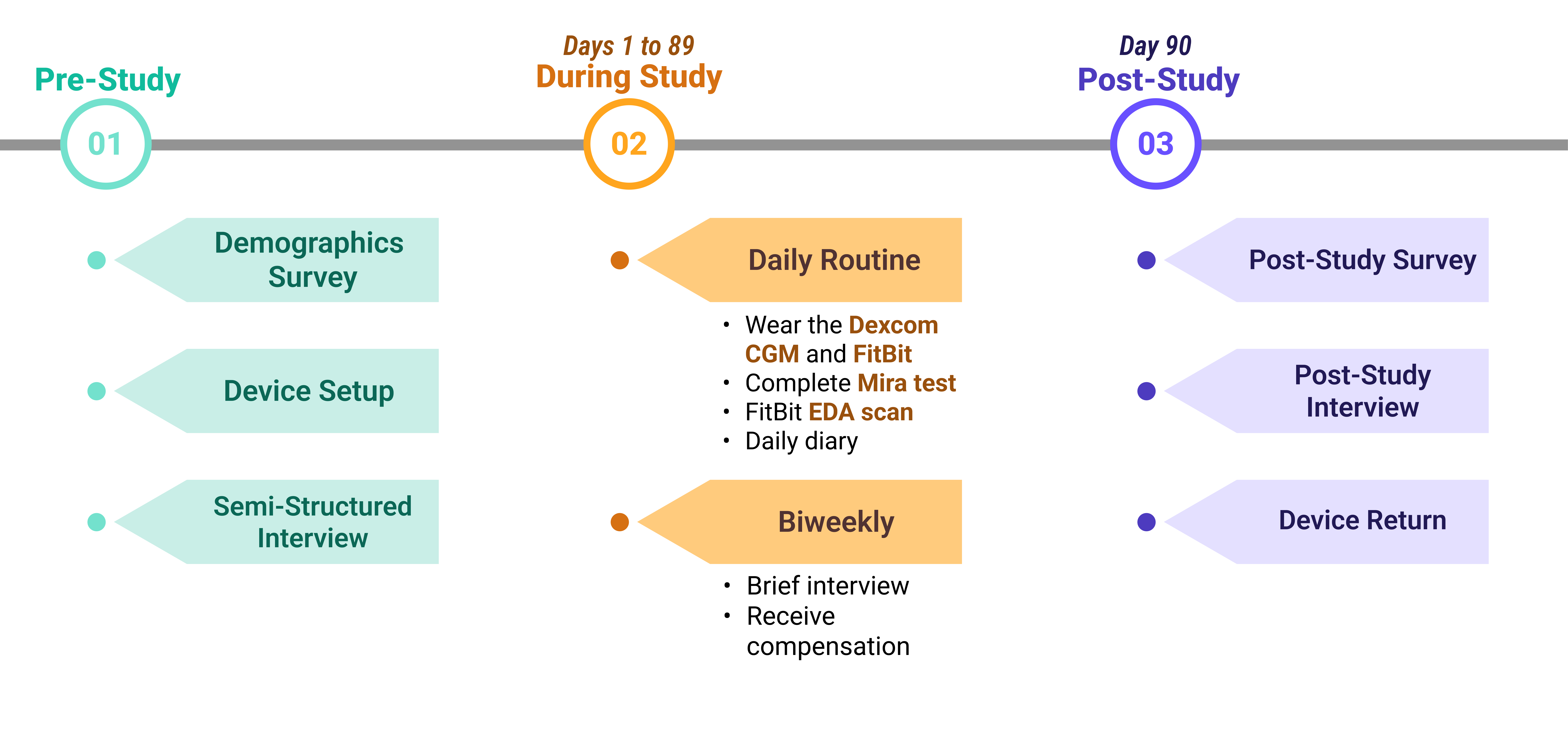}
    \caption{The protocol that participants followed over the three-month data collection period. Participants engaged in a daily routine that involved using three devices and reflecting on their experiences with a daily diary. Every other week, researchers interviewed participants to review their progress and experiences. 
    }
    \Description{A graphic representing the protocol that participants followed over the three-month data collection period. Participants engaged in a daily routine that involved using three devices and reflecting on their experiences with a daily diary. Every other week, researchers interviewed participants to review their progress and experiences. }
    \label{fig:methods-timeline}
\end{figure*}

Participants were asked to collect data over the course of three months so that the study would extend beyond a single menstrual cycle.
We chose to have participants view their data through the devices' existing apps rather than a bespoke interface that we could have created to collate and synthesize the data.
This decision was made for two reasons: (1) we wanted to explore how participants would reason about their menstrual health using products that are currently available in the market, and more importantly (2) we did not want to nudge participants towards particular signal relationships based on intentional or unintentional design decisions.

The devices' apps already provided some background on the signals they collected, and the devices also came with online resources and forums if participants chose to explore them.
Therefore, we did not provide any additional background information about the collected data signals to avoid further influencing participants' perspectives on them.

Of the 50 participants who were initially enrolled in the protocol, 40 (80\%) stayed until the end of their three-month term, resulting in a median engagement period of $90 \pm 22$ days.
Most participants who dropped out early left one month before the end of their three-month term in the study.
The protocol timeline is shown in \autoref{fig:methods-timeline}, and a summary of the data they tracked is shown in \autoref{tab:methods-data}.
We describe each component of the protocol below:
\begin{table*}[!t]
    \caption{The data reported by and accessible to participants during the three-month data collection period.}
    \label{tab:methods-data}
    \Description{Physiological and self-reported data collected such as sleep, heart rate, oxygen, and exercise collected by Fitbit sense; glucose collected by Dexcom G6; menstrual hormones collected by Mira hormone starter kit; and onset of menstruation characteristics through daily diaries.}
    \begin{tabular}{ l c }
    \toprule
      \textbf{\space  } & \textbf{Fitbit Sense} \\
      \hline
          \textbf{Sleep (daily)} & Sleep score, sleep duration, sleep phase statistics\\ \hline
          \textbf{Heart Rate (daily and hourly)} & Resting, heart rate variability\\ \hline
          \textbf{Oxygen (daily and hourly)} & SpO$_2$, VO$_2$ max, respiratory rate\\ \hline
          \textbf{Exercise (daily and hourly)} & Non-stationary minutes throughout day, time in heart rate zones \\ & calories, exercise, distance, 
            steps\\ \hline
          \textbf{Mindfulness (daily)} & Electrodermal activity, readiness score \\ \hline
          \textbf{Other (daily and hourly)} & Altitude, body temperature, wrist temperature\\
      \midrule\midrule
        \textbf{\space} & \textbf{Dexcom G6} \\
    \hline
          \textbf{Glucose (hourly)} & Continuous glucose levels (mmol/L)\\
    \midrule\midrule
      \textbf{\space} & \textbf{Mira Hormone Starter Kit} \\
    \hline
          \textbf{Menstrual Hormones (daily)} & Luteinizing hormone level (mIU/ml), \\ & Estrogen level (ng/ml)\\
    \midrule\midrule
      \textbf{\space} & \textbf{Daily Diary} \\
    \hline
          \textbf{Onset of Menstruation Characteristics} & Menstruating day, flow, color, texture\\ \hline \textbf{Symptoms} & Appetite, exercise level, libido, breakouts/acne, headaches, hot flashes,  \\ & cramps, tender breasts, fatigue, sleep issues, mood swings, stress\\ & 
          food cravings, indigestion, bloating, arousal and valence, \\
    \bottomrule
    \end{tabular}
\end{table*}
\begin{figure}
    \centering
    \includegraphics[width =0.7\textwidth]{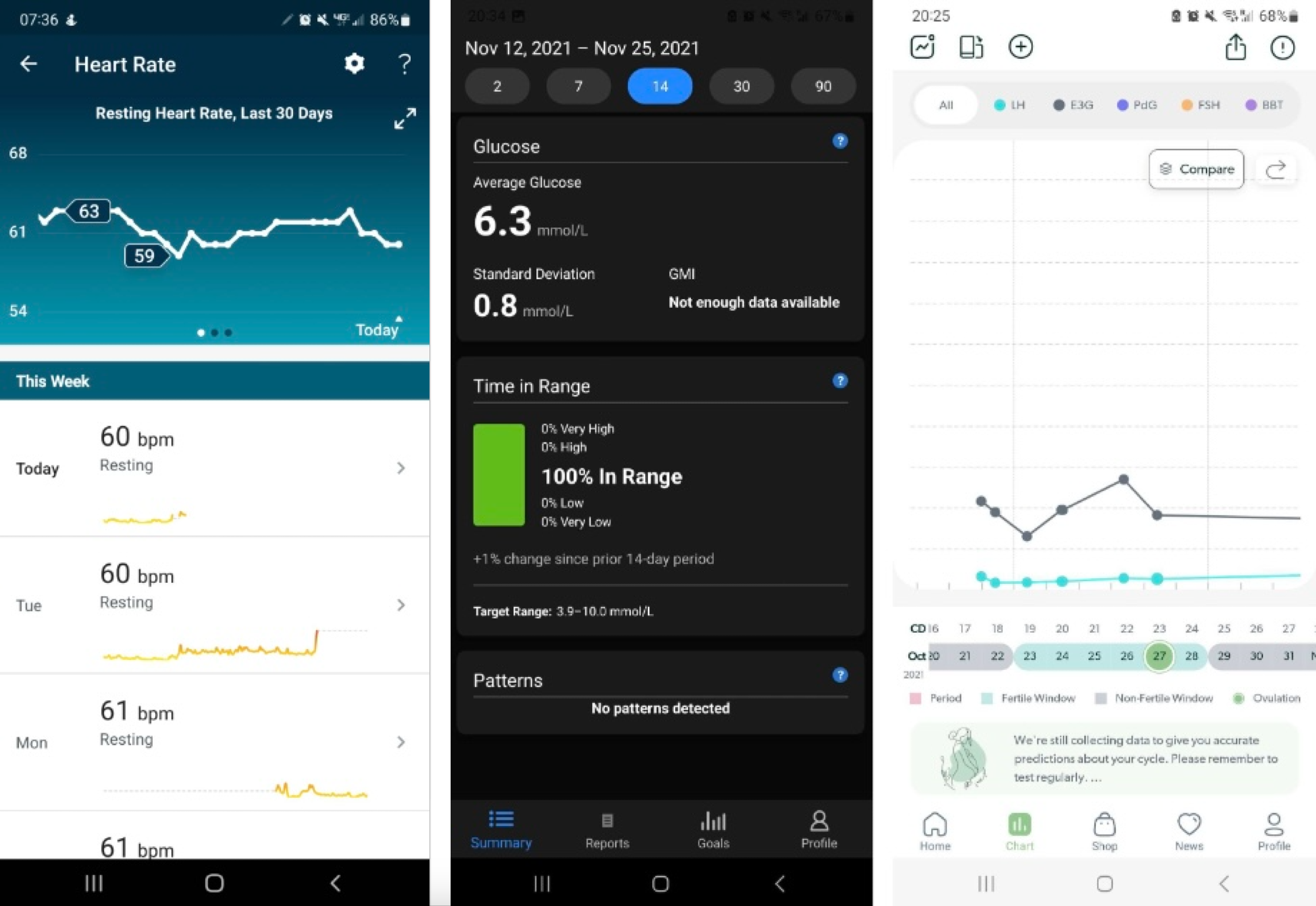}
    \caption{(left) Fitbit's companion smartphone app page for examining trends in heart rate. 
    The app has a separate page for each data stream to support in-depth analysis of a specific signal.
    (middle) Dexcom's companion smartphone app shows statistics of the user's glucose level over a selected timeframe. 
    (right) Mira's companion smartphone app shows the user's LH and E3G levels over time. At the bottom, the app provides predictions for the start and end dates of different menstrual phases (ovulation in pink, menstruation in blue).}
    \Description{(left) Fitbit's companion smartphone app page for examining trends in heart rate. 
    The app has a separate page for each data stream to support in-depth analysis of a specific signal.
    (middle) Dexcom's companion smartphone app shows statistics of the user's glucose level over a selected timeframe. 
    (right) Mira's companion smartphone app shows the user's LH and E3G levels over time. At the bottom, the app provides predictions for the start and end dates of different menstrual phases (ovulation in pink, menstruation in blue).}
    \label{fig:devicescreens}
\end{figure}

\subsubsection{Pre-Study Interview}
Before the study began, participants completed a demographics form and a 30-minute semi-structured interview. 
Participants were asked to elaborate on their medical history, particularly as it related to their menstruation, and any prior experiences they had with menstrual tracking.
Example prompts included the following: “Describe your menstrual experience over the past few years”; “Describe your health tracking history”; and “Why do you wish to track your menstrual cycle?”.

\subsubsection{Continuous Data Collection}
During the onboarding process, participants installed the companion smartphone apps for the Fitbit, Dexcom, and Mira devices in order to view the corresponding data.
A subset of the screens they saw through the apps is shown in \autoref{fig:devicescreens}.
Participants were instructed to charge their Fitbit every day for short periods of time (e.g., during showers) and to replace the CGM sensor modules every 10 days when they expired.

\subsubsection{Daily Data Collection}
To emulate current practices for menstrual tracking, participants completed a daily diary through a custom smartphone app.
The diary entries asked participants to record their affective state according to arousal and valence, the characteristics of their menstruation (e.g., flow amount and color), the magnitude of symptoms associated with menstruation (e.g., fatigue, cramps), and any special events that occurred during the day.

Participants measured their LH and E3G levels each morning using the Mira hormone analyzer, which measures hormone levels via single-use, disposable urine test wands that activate after a 16-minute waiting period.
The test requires users to refrain from drinking liquids two hours prior to collecting a sample. 
For this reason, participants were encouraged to complete the test shortly after awakening.
The Mira smartphone app not only showed participants their hormone levels but also predictions for the start and end dates of their menstrual phases.

\subsubsection{Biweekly Check-Ins}
\label{sec:methods-biweekly}
Every two weeks, researchers contacted participants virtually using their preferred form of communication (e.g., email, Discord) to gather any reactions they had to their recent self-tracking experiences.
Participants were encouraged to express their thoughts organically, devoid of any specific prompts from the researchers.
These reactions included but were not limited to: comparisons between devices, recollections of how devices were integrated into their daily lives, and responses that they had to their data.
Participants were also given financial compensation on these dates to encourage continuous study adherence.
They received \$5 CAD for each day of complete data collection, which entailed wearing both the Fitbit and Dexcom devices for at least 18 hours and completing the manual data collection procedures.

\subsubsection{Post-Study Interviews}
\label{sec:post-study-interviews}
Upon finishing the protocol, the 40 remaining participants completed a 60-minute semi-structured interview to reflect on their experiences with the technologies involved in the study.
They were asked to elaborate on each data stream individually as well as any relationships they may have observed between data streams.
Participants were also asked about the impact interpreting these data sources had on their daily routines and understanding of menstrual health.
Example prompts included the following: “What data were you interested in and why?”; “Describe your tracking experience throughout the study”; and “Did your perspective on your menstrual experience change?”.

\subsection{Analysis}
\label{sec:analysis}
The 50 pre-study, 187 biweekly, and 40 post-study interviews were all recorded and transcribed.
Three researchers independently analyzed all open-ended daily survey questions and interview transcripts using open coding on each response.
The researchers then discussed code conflicts and missing codes until a consensus was reached.
Common patterns from the consolidated list of codes were aggregated using thematic analysis~\cite{braun2006using}, resulting in 16 themes such as "expanded menstrual health perceptions", “using data to validate preconceived notions”, and “uncertainty about irregularity”.
Throughout the paper, we attribute participant responses using the notation PX.

\subsection{Ethical Considerations}
The collection of personal health data, particularly when stored by third-party entities, can impose privacy compromises on research participants.
To alleviate some of these concerns, we created accounts for each participant using unique alphanumeric identifiers.
Participants were given the credentials to these accounts so that they could access their own data.
They were instructed not to add any personally identifiable information to these accounts, but they were free to add other information to improve the accuracy of the device's measurements (e.g., weight information for Fitbit's calorie burn counter).
Despite using bespoke device accounts for research, participants were still informed of all the privacy risks associated with our data collection methods.
Participants who felt uncomfortable with the handling of their data were allowed to withdraw from our study, in which case they could have requested that we delete the accounts assigned to them.
No such occurrences happened during our study.

Participants were also not restricted from using their existing tools to avoid disrupting their other health-monitoring practices.
For example, some participants continued using their own period-tracking smartphone apps to receive predictions for menstruation onset, while others continued using their fitness monitors in parallel to maintain their step count targets and streaks.
Since these methods rarely provided additional information or visualizations that were not covered by our devices, they did not interfere with the aims of our research.
Participants were also allowed to download their data after leaving the protocol in case they wished to transfer it over to their other tracking tools.

Finally, we recognize that giving people additional data about their health can have negative consequences on how they regard their body image, especially when that data is associated with an intimate topic like menstrual health~\cite{keyes2020reimagining, tuli2022rethinking}.
Although we stopped short of imposing our own views and opinions on the participants, we took multiple measures throughout our work to mitigate potentially negative thought patterns.
During enrollment, we explained the motivation of our work towards understanding the limitations of existing menstrual trackers.
We also informed participants that they were free to skip any questions or stop any interviews if they felt uncomfortable with the topics being discussed. 
Interviewers were instructed to raise any potential concerns expressed by participants to the broader research team, which included an expert in psychology and gendered health; however, no such concerns emerged during the study.

To avoid affirming any normalized societal stigmas, we encouraged participants to reflect on the societal views that were associated with their perception of their bodies \cite{tuli2022rethinking}.
For example, when participants perceived menstrual irregularities, we followed up with questions on what irregularity meant to them, why they perceived irregularity, and the positive and negative associations they had around irregularity.
Whenever participants expressed concern about their menstrual irregularity or any other aspect of their data, we advised them to seek the opinion of a medical professional.
\section{Findings}
\label{sec:find}
\subsection{RQ1: Reassessing Menstrual Health Preconceptions with Multimodal Menstrual Tracking}
\subsubsection{Confirming Hypotheses About Inaccessible Menstrual Health Data}
\label{sec:rq1-confirm}
Prior to the study, some participants suspected that aspects of their menstrual health might be related to signals they had previously found difficult to track.
While some participants had collected heart rate (N=12) and sleep (N=19) data in the past, none of them had ever tracked glucose or hormonal data. 
Participants were excited to confirm hypotheses related to the latter set of data since manual recording relevant information in current menstrual trackers was burdensome.
For instance, using the Dexcom device to passively collect glucose data was viewed as a more convenient alternative to logging meals.
Participants were especially enthusiastic about collecting hormone data since they readily understood how that data related to their menstrual health.

Participants acknowledged that these previously inaccessible data streams could eventually be used to enhance the accuracy of their menstruation onset predictions.
\italquote{My current tracker is not very accurate since the only thing it takes into consideration are the dates of my period.}{P28}
\italquote{I wish I could see my hormone levels, but I know that is not easy to track without further testing. Also, I would love an app that can have my other health information to give an in-depth tracking of my menstruation.}{P2}
P16 had previously been told by her parents that stress might affect the timing of her period but had not thoroughly explored this connection. 
She struggled to directly link her stress levels to changes in her menstrual cycle due to the burden of manual entry in her current menstrual tracking app. 
To establish a relationship between stress and timing, she felt that she had to remember to manually record self-reported stress levels near menstruation for multiple cycles.
During the study, she realized that Fitbit would provide her with information about her stress levels throughout her menstrual cycle, making the investigation process more manageable. 
When her period arrived later than usual and she observed heightened stress levels in the few days prior, she felt that she validated this connection and wanted to explore it even further during future cycles.

P7 mentioned a longstanding belief that they were more likely to experience depressive episodes at certain points during their menstrual cycle. 
However, they did not feel inclined to track their mood throughout their cycle as they did not perceive a direct alignment between mood changes and specific days in their menstrual cycle (e.g., the first day of menstrual bleeding). 
They hoped that examining hormones would allow them to better understand the cyclical nature of their emotional fluctuations. 
After the study, P7 observed that fluctuations in their mood seemed to coincide with changes in their hormone levels rather than specific days in their menstrual cycle. 
\italquote{I'm looking at the hormone data and I'm like, “Okay, when I'm starting to notice a shift in the hormones going up and I'm feeling horrible \dots~This is what's happening within the week.”}{P7}

\subsubsection{Reflecting on Menstrual Cycle Regularity}
\label{sec:rq1-reflect}
Fourteen participants reported in the pre-study demographics form that they had irregular periods. 
These individuals and others often sought to compare their data relative to what they deemed “normal”, either with respect to their own past experiences or the experiences of others. 

Despite being taught to expect regular menstrual cycles from online resources and health education classes, they struggled to define menstrual irregularity.
They grappled with establishing standard thresholds to define irregularity according to characteristics that were easy for them to track such as the interval between days of menstrual flow and the duration of menstruation.
While P6 thought that an average of 20 days between days of flow was too short, P3 and P13 thought that averages of 21 and 18 days were standard.
P16 spoke about how she was taught that a normal period is between 28 days and 30 days, but participants like P47 had been told that 35 days was completely typical.
\italquote{You know how at the start of your period there might be symptoms, and they are more intense, and then they just kind of mellow out towards the end? I don't know if that's pretty common?}{P45}

Participants also noted that their perceived regularity status changed over time, making it more challenging for them to define irregularity.
P9 contrasted the predictability of her menstruation dates and flow color over the last five years relative to the past six months.
She noted that her period became more regular since the interval between days of menstrual bleeding had increased and her flow color had brightened.
On the other hand, P40 believed that her menstrual cycle had become irregular in the past year since her cycle began to last longer than it did in previous years. 

Without a clear definition of regularity in observable dimensions and a constant impression that there exists a standard menstrual cycle, participants felt burdened with worry and stress as they operated on the assumption that their experiences were atypical.
Supplying them with unconventional menstrual health data motivated them to re-evaluate their definition of irregularity.
Participants found it especially intriguing to track physiological changes, as they believed that drawing connections between those changes and events in their menstrual cycle would validate the flux they experienced in their bodies.
\italquote{Seeing the [physiological] data made me feel they are more real, that they happened and impacted my body.}{P7}
\italquote{I think everything is due to my hormones, but I don't know how they affect my symptoms.}{P42}
Moments of elevated stress like exams, travel, and periods of high workload were all described as notable and uncontrollable circumstances that impacted participants' menstrual health.
Therefore, individuals like P16 and P45 paid close attention to their Fitbit stress scores to validate assumed correlations between high-stress days and events in their menstrual cycle.
Even if participants did not experience any consequences as a result of drastic routine changes, they expected to see those changes reflected in their sensor data.
\italquote{My sleeping schedule's a bit muddled up right now, mostly because of Ramadan and stuff like that. Like you usually wake up a couple of hours before sunrise and then go back to sleep, so \dots~I was interested in knowing what the Fitbit [Sleep] had to say about that.}{P18}
When their expectations were not reflected in their sensor data, participants grew frustrated because they felt that they had exhausted their options in exploring potential explanations for their perceived cycle irregularity.

\subsection{RQ2: Changes in Routine Behaviors Due to Multimodal Menstrual Tracking}
\subsubsection{Reassessing Current Menstrual Tracking Practices}
\label{sec:rq2-reassess}
Prior to the study, several participants (N=19) stated that they only used their menstrual trackers to log dates of menstruation.
Twelve participants explicitly told us that their past menstrual tracking practices were centered around taking note of when menstruation began.
\italquote{It's hard to remember to use it [current menstrual tracker] regularly, I only use it to track the dates of my period.}{P20}
Participants were not keen on tracking symptoms during other menstrual cycle phases, as they neither perceived a need to do so nor knew what they could gain from tracking them.
\italquote{I pretty much only use the date and flow features on the app. I don't really track anything else because I don't feel like the app gives me feedback when I input extra things like mood.}{P44}

As the study progressed and participants engaged with dynamic physiological data, the fluctuating nature of step count, sleep, and exercise data captured their interest. 
They began to reflect on their self-reported symptom data throughout their cycles, not just during days of menstruation, to investigate whether that data displayed variability comparable to that of the dynamic physiological signals.
For example, P18 conveyed that she examined her activity levels more often because she cared about her exercise routine.
While checking her activity levels, she became increasingly curious about how changes in her exercise routine might have been impacting her symptoms.
By examining this data across days and weeks, she reported having a heightened awareness of the potential connections between her mood and activity levels.

\subsubsection{Planning Schedules Around the Menstrual Cycle}
Most participants (N=42) had been tracking their menstrual cycle to avoid scheduling important events during days with menstrual bleeding or elevated menstrual symptoms.
Providing participants with multimodal menstrual tracking augmented the way they planned their daily lives.
Beyond looking at their self-reported data to decide when adjustments needed to be made, many (N=32) adjusted their routines to accommodate fluctuations in their physiological data in anticipation of potential menstrual symptoms.
For example, P18 began to recognize which events were more likely to increase her physiological stress.
Upon realizing that heightened stress also negatively impacted her mood, she rearranged her schedule so that low-stress activities were scheduled closer to menstruation.
Others (N=22) utilized the daily hormone data to identify phases in their menstrual cycle beyond menstruation during which they preferred to engage in or refrain from certain activities.
After observing emotional differences across phases of her menstrual cycle, P45 began proactively planning stress management activities in anticipation of phases during which she felt her mood was lower.
Meanwhile, P14 avoided engaging in sexual activity during days when the Mira app indicated a higher likelihood of conception.

The majority of participants (N=30) indicated a change in their daily routines with the aim of improving their overall health. 
Their motivation stemmed from uncertainty regarding whether all the signals were specifically linked to their menstrual health. 
Consequently, they were inspired to make lifestyle adjustments in the hopes of improving their menstrual health and their general well-being.
\italquote{I have been eating better and maintaining my blood sugar. I find that some symptoms like cravings and headaches naturally went away.}{P25}
\italquote{I tried to sleep more after days my sleep score wasn't good, exercise more after the days I had a low step count, and eat healthier during the days I had a low calorie output.}{P22}

\subsubsection{Sparking Conversations Around Menstrual Health}
Giving participants access to unconventional menstrual health signals provided them with opportunities to seek guidance and advice. 
Seventeen participants shared their experiences in the study with friends or family.
Eleven participants mentioned seeking advice from these individuals, especially from others who were participating in the study.
Some even enrolled together as roommates and siblings and were excited to discuss their data with one another.

Many participants encountered challenges initiating direct conversations about menstrual bleeding, as they were concerned that people might feel uncomfortable discussing the topic.
\italquote{My dad has been looking after me for 23 years but gets very uncomfortable with this topic. So like, for me, it was really weird to just bring my period up to people because I don't know how they will receive it.}{P41}
Participants felt more at ease using unconventional menstrual health signals to prompt conversations about both their menstrual health and other health-related topics.
Twelve participants had conversations about their hormone levels, ten participants initiated conversations about their glucose data, and seven participants regularly compared their sleep and readiness scores with other participants.
\italquote{[Another participant] and I really enjoy comparing our sleep scores together. We then check in with each other about our hormone levels and how it’s making us feel. It gives us a lot of good insight into how the other person is doing on a day-to-day basis.}{P47}

When participants had similar data to one another, they felt like they were part of a majority and thus more “normal”; if their data differed, they started to feel anxious and concerned about a possible medical issue.
\italquote{[Talking about another participant] She's actually worried now about her fertility because of Mira. She's worried that her estrogen doesn't spike the same way mine does, so I think it's caused a lot of confusion on her part.}{P10}
As these worries grew, some participants considered gathering unconventional menstrual health data to guide discussions with medical professionals.
For instance, P24 had concerns about her energy levels throughout her menstrual cycle, so she brought her glucose and hormone data to her doctor to validate hypothesized relationships.
Likewise, P25 shared her glucose data with her doctor to understand whether low blood sugar levels could explain diminished physical activity levels.
\subsection{RQ3: Increased Menstrual Health Understanding Due to Multimodal Menstrual Tracking}
\subsubsection{Viewing Menstrual Health Beyond Periods of Bleeding}
\label{sec:rq3-viewing}
Participants were pleasantly surprised that the Mira device was able to identify different phases of the menstrual cycle using hormone data.
\italquote{Prior to the study, I was already kind of tracking my menstrual cycle \dots~But I had never seen my ovulation phase or the hormones go up or down before. It was new for me and really interesting to see that.}{P27}
Even though some participants had been taught about different phases of the menstrual cycle during a sexual education course, being reminded of these phases led to a significant shift in how they approached their data.
\italquote{[The Mira device] just sort of broke it down into how many days each phase was for me, and I liked how it broke it down into those phases. With some of the symptoms that I'm experiencing, it also helped me understand myself and my cycle better.}{P14}

Participants observed cyclical signal fluctuations that manifested throughout the menstrual cycle and not just during days of menstruation.
Expanding their temporal purview of the menstrual cycle empowered many individuals to connect different phases and symptoms, yielding explanations for past experiences that previously seemed unrelated to their menstrual health.
In fact, eight participants reported discovering that symptoms they had previously associated exclusively with menstruation occurred during other phases, such as P9's observation that her acne became more prominent near ovulation.

Other participants were surprised to find that their symptom severity levels were higher during phases beyond menstruation.
For example, menstrual trackers often assume that cramps are most prominent during menstruation~\cite{epstein2017examining}, and 19 participants confirmed this assumption by giving the highest average severity ratings for cramps during that phase of their cycle; however, 17 participants experienced the highest cramp severity during ovulation, while 11 participants experienced the highest cramp severity during their luteal phase.
Nevertheless, some participants were unable to discern how some signals and symptoms related to their menstrual health if they were only notable during non-menstruation phases. 
\italquote{I did have a bit of confusion because I couldn't tell whether it was bloating pain or if it was cramping [when I was not menstruating].}{P26}

\subsubsection{Synchronizing Hypothesis Testing With the Menstrual Cycle}
Participants often utilized a trial-and-error approach to observe and interpret the relationship between various physiological signals and their menstrual health. 
They anticipated refining their interpretations as each cycle passed, engaging in a cyclical process of learning from their data.
Some participants were particularly reflective around times when they noticed significant changes in signals. 
For example, P48 consistently noted significant hormone spikes early in the study, which led her to realize that some spikes had a consistent temporal offset to her menstrual bleeding. 
Consequently, she focused less on menstrual timing predictions during subsequent cycles until she observed hormone spikes.
Similarly, P9 was excited to discover how significant glucose changes were correlated with her menstrual health.
\italquote{Whenever I thought there was supposed to be a change, that's when I would be curious. Like when I noticed more changes in glucose when I was lower in estrogen, I thought that was strange. So I Googled that and they said that estrogen regulates insulin.}{P9}

However, participants faced challenges when attempting to confirm relationships between physiological signals and menstrual cycle symptoms. 
The complex and varied contextual factors, such as interwoven physiological processes and both inter- and intra-cycle variances, complicated their efforts to infer how their sensor data may be related to their menstrual symptoms. 
For example, P9 observed that her resting heart rate fluctuated every few weeks but cited a lack of medical understanding when trying to determine if that pattern was related to her menstruation.
While P15 was more confident about discerning relationships within her data, she was uncertain about the number of cycles she needed to review before validating whether she was observing a coincidence or a trend.
\italquote{I had four cycles during the study, and I think the last two were kind of similar \dots~but the other two were different. I was like, "I don't know what counts as a trend" \dots~I did things [for other parts of the cycle], but when should it [hormones] go up or down? I don't know.}{P15}

\subsubsection{Understanding Menstrual Health Through Trustworthy Data Collection Methods}
\label{sec:q3-2}
For almost all participants (N=46), this study marked their first experience interacting with their own blood or urine for data collection, with blood being used by the Dexcom device and urine being used by the Mira device. 
As they reflected on which data sources influenced their menstrual health understanding more, they expressed a higher level of trust in devices that relied on body fluids compared to other forms of data collection.
They perceived these devices as closer to clinical tests, even if they applied an additional layer of post-processing for smoothing data or generating labels (e.g., Mira's phase predictions).
\italquote{I think that Mira is the most relevant period-checking app \dots~because it only focuses on hormones.}{P16}
\italquote{I know it [current menstrual tracker] makes predictions based on dates of your previous cycles, and then it uses that to make a prediction of the next cycle. But I like that, with Mira, it combines both [dates and urine test].}{P18}
This preference for seemingly clinical data lies in stark contrast to devices that participants felt heavily relied on signal processing or machine learning, namely Fitbit's algorithmic predictions for sleep and stress. 
\italquote{I think previous algorithms mostly base it off of population data at first, like what you typically see. But it sort of has this dimension missing \dots~It needs an extra layer of data that can refine its algorithms.}{P18}

Although many participants acknowledged the benefits of using the Dexcom and Mira devices, they were reluctant to continue using them after the study due to the devices' target audiences.
Ten participants commented that the Dexcom device was intended for diabetics and therefore felt that it was not designed for them as non-diabetics.
Similarly, eight participants said that they would only use the Mira device if they wished to get pregnant in the future.
\italquote{I wouldn't [use Mira or Dexcom again] since I don't have diabetes or hormone problems. If I were trying to get pregnant, I'd use Mira until I became pregnant and then I would stop.}{P10}

\section{Discussion}
\label{sec:discussion}
Many of today's menstrual trackers are designed to monitor the date and duration of people's menstruation~\cite{epstein2017examining, zwingerman2020critical}, thereby shaping how they interpret their menstrual health~\cite{tuli2022rethinking}.

Our study shows that giving people access to unconventional menstrual health signals like hormone, glucose, and stress data can push them to reflect on their menstrual health in other ways.
Below, we first describe how participants shifted towards a more expanded view of menstrual health.
We then discuss the benefits they received from this perspective shift.

\subsection{Changes in Current Menstrual Health Practices}
During our study, many participants experienced an evolution in their understanding of menstrual health, particularly with respect to consistency and normality.
While their initial focus was primarily on bleeding during menstruation, participants began to acknowledge the relevance of other menstrual cycle phases.
Participants also explored unconventional menstrual signals to quantify underlying mechanisms that influence their menstrual health, prompting them to consider physiological measurements rather than exclusively observable symptoms.

\subsubsection{Exploring Menstrual Cycle Phases Beyond Menstruation}
\label{sec:dis-phases}
Participants exhibited high confidence in Mira's predictions because they were derived from bodily fluids, unlike the algorithmic stress and sleep scores provided by their Fitbit devices.
As participants examined fluctuations in their hormone levels alongside Mira's menstrual phase labels, they came to recognize the potential significance of the intervals between menstruation. 
Acknowledging the follicular, ovulation, and luteal phases in addition to menstruation prompted them to consider every day as part of a cyclical pattern rather than fixating strictly on the days with menstrual bleeding.
Participants believed that the regulation of various physiological processes was dictated by hormones, aligning with extensive medical literature that links hormonal fluctuations to diverse physiological processes throughout the menstrual cycle~\cite{baker2020temperature, lin2023blood, brar2015effect}.
Consequently, some participants leveraged their newfound or renewed knowledge of menstrual phases to not only identify additional milestones in their menstrual cycle but also to associate physiological changes with those milestones.
This allowed them to distinguish between fluctuations that may be relevant to their menstrual health and those that were not.

~\\\noindent
\textbf{Design Implications:}

Future menstrual trackers should include some way of presenting phases of the menstrual cycle when displaying historical or predicted menstrual data. 
This approach would allow users to recognize that the interval between consecutive instances of menstruation may have relevance to their menstrual health.
Given that phases beyond menstruation are challenging for participants to identify based solely on observable symptoms, designers should consider integrating signals that are able to determine these phases automatically.
While using hormone data remains the gold standard for teasing apart menstrual phases, collecting such data remains financially and physically burdensome, and our participants did not foresee themselves using the Mira device beyond the study because they did not associate themselves with its target audience.
Since self-reported dates of menstruation are not enough for accurate phase predictions~\cite{epstein2017examining, zwingerman2020critical, fox2020monitoring}, there have been growing efforts to apply machine learning on physiological data to this end~\cite{shilaih2017wearable, shilaih2018wearable, goodale2019wearable, hayano2021prediction}.
Still, the outputs of these models are typically limited to a subset or semantic grouping of menstrual phases (e.g., menstruation vs. non-menstruation).
Our work highlights the importance of models that rely on accessible and convenient data to generate fine-grained menstrual phase predictions.

Although identifying fine-grained menstrual phases can provide benefits to users, doing so also comes with legal considerations with respect to privacy~\cite{dong2022privacy}.
Rather than relying on sensitive and easily interpretable hormone data, leveraging multimodal physiological data for phase prediction may mitigate some of these concerns.
Nevertheless, menstrual trackers should offer users the option to opt out of menstrual phase prediction to respect their privacy and autonomy.
Designers might also consider enabling users to track their menstrual phase independently without needing to store hormone data within the application.

\subsubsection{Expanding Current Menstrual Tracking Using Passive Sensing}
As participants grew to appreciate the significance of each menstrual cycle phase, they were eager to integrate automatically recorded physiological signals into their tracking practices.
By having the Fitbit and Dexcom devices continuously collect data on their behalf, participants were able to examine facets of their health across their entire menstrual cycles, not just during the days leading to menstruation.
Reflecting on associations across signals engendered increased awareness of well-being, mirroring the benefits observed in other forms of multimodal health tracking~\cite{bentley2013health,choe2017understanding, kim2017omnitrack}.

Similar to participants in past studies outside of menstrual health~\cite{bentley2013health,choe2014understanding,choe2017understanding}, our participants were more keen on analyzing signals that exhibited significant or unexpected changes rather than signals that were consistent or predictable.
Their motivations for menstrual tracking often revolved around the timing of observable symptoms, leading them to temporally align times of flux with the menstrual cycle. 
Giving participants the ability to identify cycle phases beyond menstruation gave them yet another dimension with which to perform this alignment, thereby yielding additional markers for triangulating their experiences across multiple cycles.
This resulted in a renewed motivation to manually self-report symptoms as participants sought to understand if times of flux during certain various menstrual phases resulted in menstrual symptoms.

~\\\noindent
\textbf{Design Implications:}
The length of a person's menstrual cycle can naturally vary over time~\cite{brantelid2014menstruation}, requiring individuals to manually record data consistently and daily if they want to be prepared for semi-unpredictable moments of interest or to compare trends between cycles.
The introduction of passively tracked menstrual signals may offer participants relief from the burden of manual tracking while also helping them determine when to focus or defocus on their data.
A modified menstrual tracking practice could involve consistent tracking of passive signals until a major deviation in a signal is predicted or observed, prompting users to engage in more detailed tracking like symptom logging.

At the same time, it is important to note that the absence of significant changes in a signal within a menstrual cycle may also hold relevance for some users. 
For instance, individuals trying to conceive may deem menstrual hormone spikes during the ovulation phase as meaningful indicators of successful ovulation, whereas non-chronic anovulatory cycles with subtler hormonal changes may be less relevant to those not aiming to conceive~\cite{reed2015normal, prior2015ovulation}. 
Therefore, individuals' motivations for menstrual tracking should be considered when suggesting different menstrual tracking practices, and future studies should explore how these practices impact users' perception of cycle regularity and signal relationships over time.

\subsection{Benefits of Incorporating Unconventional Menstrual Health Signals}
\label{sec:dis-unconventional}
The bidirectional relationship between menstrual health and overall well-being is widely acknowledged among researchers~\cite{pichon2021messiness, goodale2019wearable, shilaih2017pulse, yu2022tracking, luo2020detection, grant2020ultradian}, yet it is not reflected by most menstrual tracking applications.
Incorporating unconventional menstrual health data, particularly signals that are less commonly associated with menstrual health directly, could help people better appreciate this relationship.
Our findings lead us to the notion that there exists a continuum of how people view their menstrual health.
At one end of the spectrum is a \textit{reductionist}\footnote{We use the term `reductionist' not as a disparaging term, but rather to convey due to intentional or unintentional simplification of the nuances of menstrual health.} view wherein one's scope of menstrual health is limited to conventional menstrual health signals.
On the other end of the spectrum is a \textit{holistic} view wherein one's understanding of menstrual health accounts for all aspects of health, which includes relationships between conventional and unconventional menstrual health signals.

\begin{figure}
    \centering
    \includegraphics[width=0.8\linewidth]{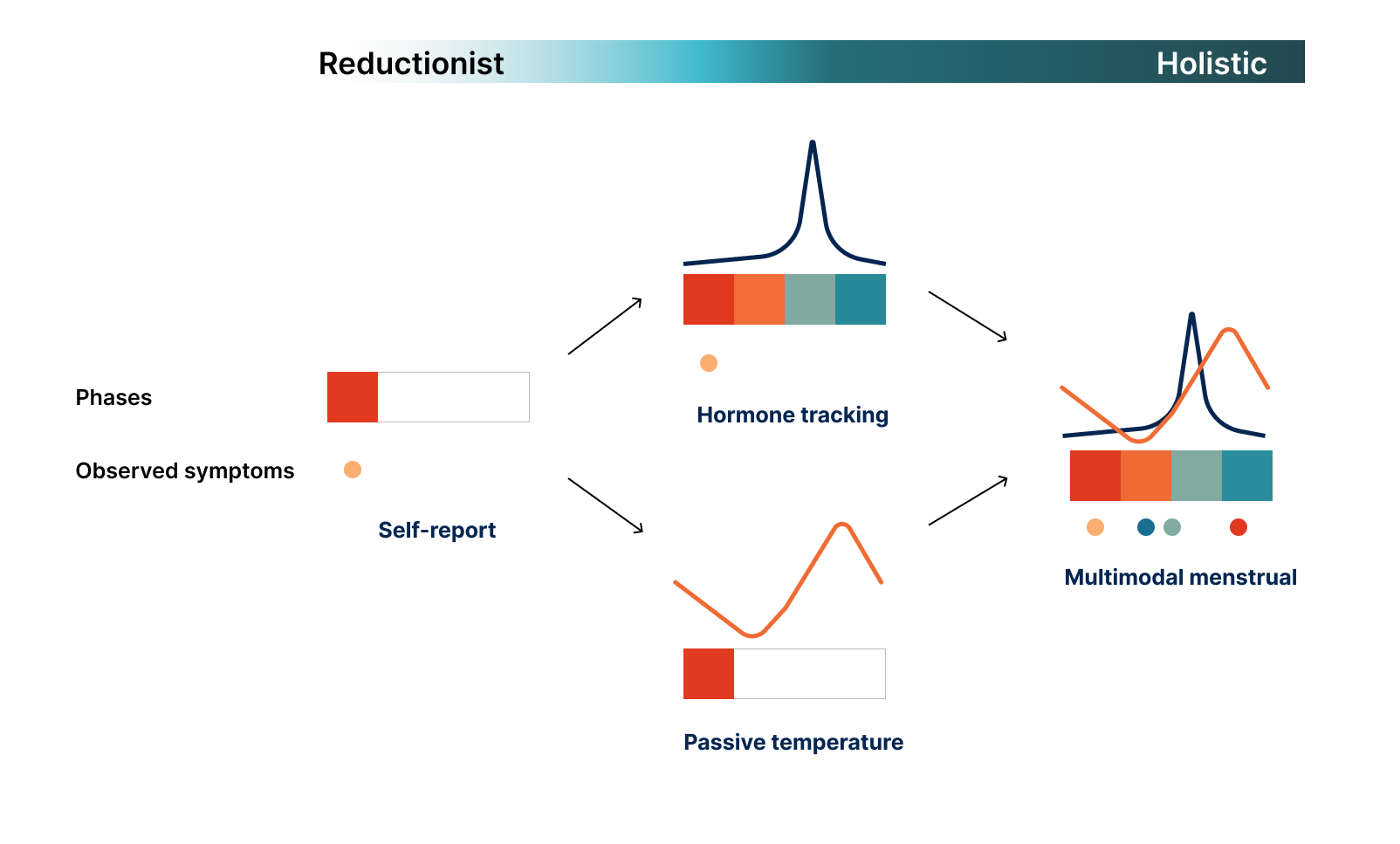}
    \caption{By interacting with hormone data, participants expanded their purview of the menstrual cycle beyond menstruation. They discovered that fluctuations in passively tracked signals aligned with these other phases, leading them to identify events and trends that may be linked with other menstrual phases or their cycle as a whole. This motivated participants to log more self-reported data to better understand observed symptoms.}
    \Description{The figure visualizes a spectrum between reductionist to holistic approaches to understanding the menstrual cycle. The left side of the figure represents a reductionist view, showing a focus on self-reported symptoms during menstruation. The middle of the figure shows how hormone tracking can lead participants to recognize the relevance of data from all menstrual cycle phases while passive tracking could lead them to collect non-self-reported data across the menstrual cycle. The right side represents multimodal menstrual tracking where both hormone tracking and passive temperature monitoring are done at the same time. These methods can lead to a holistic understanding of the menstrual cycle where multiple phases, symptoms, and physiological data are observed and correlated.}
    \label{fig:expandmenstrual}
\end{figure}

As illustrated in \autoref{fig:expandmenstrual}, individuals do not take a fixed position along this continuum but rather fluctuate depending on their lived experience and information-gathering needs.
For example, participants who only care about knowing the timing of menstrual bleeding may be satisfied with a reductionist view of menstrual health. 
Participants in our study started to move towards a holistic view by combining menstrual cycle phase tracking facilitated by hormone data with physiological tracking facilitated by passive sensing; nevertheless, most existing menstrual trackers are reductionist in how they present menstrual health~\cite{epstein2017examining, zwingerman2020critical, pichon2021messiness}.

We encourage future research to delve further into examining which features are necessary to encourage or support a more holistic perspective of menstrual health.
On the one hand, menstrual trackers could incorporate unconventional menstrual health signals alongside information about their correlation with menstrual health outcomes.
Given that many multimodal trackers assert that they are "universal"~\cite{kim2017omnitrack, bentley2013health, choe2017understanding, wu2021multimodal} without encompassing menstrual tracking features, an alternative recommendation would be for existing health trackers to integrate features dedicated to menstrual health so that they offer a more comprehensive health monitoring experience.
Providing toggleable menstrual cycle phase labels or dedicated visualizations showing cycle progression over time would allow universal health trackers to accommodate reductionist and holistic views of menstrual health.
The rest of this subsection describes opportunities and challenges in supporting a holistic view.

\subsubsection{Cultivating Positive Lifestyle Adjustments}
Past studies have found that multimodal tracking can inspire changes in behaviors and daily routines~\cite{bentley2013health,choe2017understanding,choe2014understanding}.
Unlike other personal informatics domains like physical activity and financial planning, \citet{epstein2017examining} note that individuals often perceive little to no control over their menstrual health experiences.

As participants in our study were pushed towards a holistic perspective of menstrual health, they began to identify changes in their lifestyle that they suspected could improve their menstrual health according to actionable, unconventional menstrual health signals.

Many existing menstrual trackers have been critiqued for endorsing the “treatment” and “medicalization” of menstrual health under the guise of giving people a sense of control over their periods~\cite{tuli2022rethinking, homewood2019inaction}.
Scholars have also criticized the perpetuation of dualist approaches where the mind and body are distinct and separable, leading people to believe they can “fix a natural bodily phenomenon” like menstruation~\cite{tuli2022rethinking, homewood2019inaction, homewood2020putting}.
Incorporating unconventional menstrual health data into menstrual trackers could be viewed as exacerbating these issues by motivating people to pursue new dimensions of unnecessary behavior change.
At the same time, it may also help individuals identify and change modifiable lifestyle factors that have beneficial downstream impacts on menstrual health; for example, maintaining a healthy diet can prevent dramatic changes to plasma estradiol levels~\cite{pirke1985influence} and engaging in stress management can help regulate emotion changes associated with hormone fluctuation~\cite{ossewaarde2010neural}.
Furthermore, the inclusion of unconventional menstrual health data in trackers may not only promote healthy lifestyle choices that positively affect overall and menstrual health but also educate individuals about the mutual influence of different health facets.

\subsubsection{Building Positive Associations with Menstrual Health}
Building up to a holistic view of menstrual health might also foster positive body association by linking menstrual health with other definitions of health that are not associated with as much social stigma~\cite{johnstonrobledo2011menstrual}.
In our study, we found that participants used unconventional menstrual health signals to initiate discussions about their menstrual health as they were more comfortable talking about these generic signals first.
Future menstrual trackers could further facilitate these discussions by calculating population-level statistics related to unconventional menstrual health data and then forming communities around people with similar trends~\cite{epstein2017examining}, such as diabetics or individuals who maintain a high physical activity level. 
Prior solutions for scaffolding communities and discussions around conventional menstrual health signals have often been criticized for perpetuating universal experiences and stereotypical designs, and these challenges may persist with the introduction of new signals~\cite{epstein2017examining, keyes2020reimagining}.
Still, data sharing happens naturally among individuals who practice multimodal tracking~\cite{rooksby2014personal,kim2017omnitrack}.
Initiating these communities according to commonalities in unconventional menstrual health data with the underlying pretense that menstrual health topics can be discussed may be a way of initiating conversations with less stigma.

\subsubsection{Setting Nuanced Expectations for Menstrual Cycle Regularity}
\label{sec:discussion-rq2-regularity}
Shifting closer to a holistic view of menstrual health, participants began to compare symptoms and signals across multiple time horizons (e.g., within a phase, within a cycle, across multiple cycles).
However, participants encountered inter- and intra-cycle variances within their data, causing worry when they began to compare their data with others in the study who observed different trends.
\citet{stubbs2020learning} suggest that there exist different "normals" both between and within individuals, challenging the assertion of population-level baselines embedded or implied in many current menstrual trackers~\cite{lin2022investigating, zwingerman2020critical, berkovsky2012influencing, fox2020monitoring,marchand1995feminism}. 
Emphasizing that a range of experiences may be normal can help avoid perpetuating problematic norms based on narrow population-level baselines~\cite{keyes2020reimagining}.
This shift can be complemented by incorporating educational resources that highlight the prevalence of variance in both conventional and unconventional menstrual health data.

Participants struggled to establish consistency between their cycles when they solely focused on their own data, particularly in determining the necessary number of observed cycles required to establish reliable patterns. 
Many medical sources distinguish between momentarily irregular cycles and irregular cycles in perpetuity~\cite{newton2003socioemotional, weller2002menstrual, wang2020menstrual}; a single atypical menstrual cycle is usually not a cause for concern since it could be influenced by stress or other factors~\cite{song2022factors}, yet lacking any consistency over a long time horizon warrants greater concern~\cite{wang2020menstrual}.
It is important to note that there is not a widely accepted criterion of what constitutes menstrual regularity. 
Definitions of regularity vary according to the expected interval between menstrual periods, with examples including 26–30 days, 24–32 days, and 24–35 days~\cite{weller2002menstrual}.
Definitions also vary according to the time horizon used for assessing regularity, with definitions ranging from 6–8 cycles~\cite{weller2002menstrual} to the past year~\cite{newton2003socioemotional}.
Thus, more research is needed to explore the optimal depth, frequency, and duration of data collection required to objectively assess cycle consistency. 
This might help users grasp reasonable fluctuations in physiological data over multiple cycles without overly fixating on differences between consecutive cycles, thereby alleviating concerns about momentary deviations.

\subsection{Limitations and Future Work}
\label{sec:discussion-limitations}
Our pre-study survey and subsequent interviews revealed that participants entered our study with different understandings of their menstrual health, but cultural norms and sociopolitical factors also play an important role in menstrual sensemaking~\cite{lin2022investigating}.
Our study was conducted in a single major metropolitan area in North America that skews WEIRD (Western, educated, industrialized, rich, and democratic)~\cite{henrich2010weirdest}, and the findings in this study generally reflect a Western view of feminism. 
While our participants came from a large international city with many diverse cultures, we leave further analysis or studies on this topic to future work.
We also recognize that our participant cohort was biased in other ways.
We focused our recruitment efforts on pre-menopausal adult participants, which we acknowledge may not be representative of all individuals who track their menstrual health.
Moreover, only five individuals in our study identified as non-women.
These participants did not indicate significantly different experiences related to their gender identification, but we refrain from claiming that our findings will generalize to all people who do not identify as women.
In recognition of these limitations, dedicated investigations with these populations should be conducted to better understand if there exist nuances that were not uncovered in our cohort.

Although our work vouches for health trackers that promote a holistic perspective of menstrual health, this vision could impose additional challenges on users.
A fully holistic view might entail numerous devices worn at all times to capture all health aspects and contexts.
However, leveraging multiple devices like continuous glucose monitors and hormone trackers for health tracking imposes non-trivial financial costs that we circumvented by providing equipment to our participants.
Multimodal menstrual tracking can also raise legal concerns since multiple data streams can reveal sensitive health information, such as a person's pregnancy status~\cite{dong2022privacy}.
We made the conscious decision to not integrate the various data streams given to participants so that we could understand how users would pursue holistic menstrual health views using existing technologies.
As commodity health devices become more widely available and affordable, we hope that menstrual tracker developers will investigate these important considerations in future designs.

\section{Conclusion}
In this paper, we explored people's experiences over the course of three months as they tracked conventional and unconventional menstrual health signals using physiological sensors and a daily diary.
We observed an evolution in participants' understanding of menstrual health, particularly in regard to menstrual timing, consistency, and normality.
These observations underscore the importance of features that automatically detect and present menstrual cycle phases in future menstrual trackers, bearing in mind the importance of user privacy and autonomy.
Embracing a holistic perspective of menstrual health could help people appreciate the intricate relationship between physiological signals, empowering them to make positive lifestyle adjustments and fostering healthier associations with their menstrual health.
We hope that our work leads to multimodal menstrual tracker designs that not only help people plan their schedules and prompt them to seek timely care when necessary but also combat body-related anxiety and stigmatization so that users can be more comfortable with their bodies.

\begin{acks}
 This research was funded in part by NSERC Discovery Grants \#RGPIN-2021-03457 and \#{RGPIN-2021-04268}, a Wolfond Scholarship in Wireless Information Technology, a Google PhD Fellowship, and an unrestricted gift from Google.
\end{acks}

\bibliographystyle{ACM-Reference-Format}
\bibliography{paper}

\end{document}